\documentclass[letterpaper, 10 pt, conference]{ieeeconf}  
\IEEEoverridecommandlockouts                              
\usepackage{graphicx} 
\usepackage{times} 
\usepackage{amsmath} 
\usepackage{amssymb}  
\usepackage{mathtools}
\usepackage{epstopdf}
\usepackage{subfig}
\usepackage{cite}
\usepackage[colorlinks=true]{hyperref}
\usepackage{framed}
\usepackage{multirow}

\usepackage{parskip}
\usepackage{multirow}
\usepackage{etoolbox}
\usepackage{mathptmx}
\usepackage{calrsfs}
\usepackage{lipsum}

\newtheorem{theorem}{Theorem}[section]

\newtheorem{remark}[theorem]{Remark}

\newtheorem{proposition}[theorem]{Proposition}

\newtheorem{problem}[theorem]{Problem}

\newcommand{\blockdiag}{\mbox{{\rm Block-diag}}}
\newcommand{\diag}{\mbox{$\rm Diag$}}

\newcommand{\as}{\mbox{{\rm $a.s.$}}}

\newcommand{\rn}{\mbox{$\mathbb{R}^n$}} 
\newcommand{\rnn}{\mbox{$\mathbb{R}^{n \times n}$}}
\newcommand{\rank}{\mbox{$\rm rank$}}

\newcommand{\sr}{\stackrel}

\newcommand{\rar}{\rightarrow}

\newcommand{\tri}{\sr{\triangle}{=}}

\newcommand{\be}{\begin{equation}}
\newcommand{\ee}{\end{equation}}
\newcommand{\bea}{\begin{eqnarray}}
\newcommand{\eea}{\end{eqnarray}}
\newcommand{\bes}{\begin{eqnarray*}}
\newcommand{\ees}{\end{eqnarray*}}
\newcommand{\bi}{\begin{itemize}}
\newcommand{\ei}{\end{itemize}}
\newcommand{\ben}{\begin{enumerate}}
\newcommand{\een}{\end{enumerate}}
\newcommand{\bp}{\begin{problem}}
\newcommand{\ep}{\end{problem}}
\newcommand{\hso}{\hspace{.1in}}

\title{\LARGE \bf Characterization of Conditional Independence and Weak Realizations of Multivariate Gaussian Random Variables: Applications to Networks }

\author{Charalambos D. Charalambous and Jan H. van Schuppen
\thanks{C. D. Charalambous is with the Department of Electrical Engineering, University of Cyprus, Nicosia, Cyprus. Jan H. van Schuppen is with Van Schuppen Control Research, Gouden Leeuw 143, 1103 KB Amsterdam, The Netherlands.  E-mails:
   \{chadcha@ucy.ac.cy,jan.h.van.schuppen@xs4all.nl\}}%
}

\begin{document}
\maketitle
%
%
%
\begin{abstract}
The Gray and Wyner lossy source coding  for a simple network  for sources that generate  a tuple of jointly   Gaussian random variables (RVs) $X_1 : \Omega \rightarrow {\mathbb R}^{p_1}$ and $X_2 : \Omega \rightarrow {\mathbb R}^{p_2}$, with respect to square-error distortion at the two decoders is re-examined using  (1) Hotelling's   geometric approach of Gaussian RVs-the canonical variable form,  and (2) van Putten's and van Schuppen's  parametrization of joint distributions ${\bf P}_{X_1, X_2, W}$ by Gaussian   RVs $W : \Omega \rightarrow {\mathbb R}^n $  which make  $(X_1,X_2)$ conditionally independent, and  the weak stochastic realization of $(X_1, X_2)$.\\ 
Item (2) is used  to  parametrize the lossy rate region of the Gray and Wyner source coding problem for joint decoding with mean-square error distortions ${\bf  E}\big\{||X_i-\hat{X}_i||_{{\mathbb R}^{p_i}}^2 \big\}\leq \Delta_i \in [0,\infty], i=1,2$,  by the covariance matrix of RV $W$. From this then follows Wyner's common information $C_W(X_1,X_2)$ (information definition)  is achieved by $W$ with identity covariance matrix, while a formula for Wyner's lossy common information (operational definition) is derived,  given by 
$C_{WL}(X_1,X_2)=C_W(X_1,X_2)
   =  \frac{1}{2} \sum_{j=1}^n 
        \ln
        \left(
        \frac{1+d_j}{1-d_j}
        \right),$ 
for   the distortion region  $ 0\leq \Delta_1 \leq \sum_{j=1}^n(1-d_j)$, $0\leq \Delta_2 \leq \sum_{j=1}^n(1-d_j)$, and where $1 > d_1 \geq d_2 \geq \ldots \geq d_n>0$  in $(0,1)$ are  {\em the canonical correlation coefficients} computed from the canonical variable form of the tuple $(X_1, X_2)$.   \\
The methods are  of fundamental importance  to other problems of multi-user communication, where conditional independence is imposed as a constraint. 
\end{abstract}


\section{Introduction, Main Concepts,  Literature, Main Results}
\label{sec.Intro}

In information theory and communications an important class of theoretical and practical problems is of a  multi-user nature, such as,  lossless and lossy network source coding for data compression over noiseless channels,   network channel coding  for data  transmission  over noisy channels \cite{cover:thomas:1991}, and secure communication \cite{csiszar-korner:1978}. A sub-class of  network source coding problems deals with two sources that generate at each time instant,  symbols that are  stationary memoryless,   multivariate,  and jointly Gaussian distributed, and similarly for network channel coding problems, i.e.,   Gaussian multiple access channels (MAC) with two or more multivariate correlated sources  and a multivariate output.

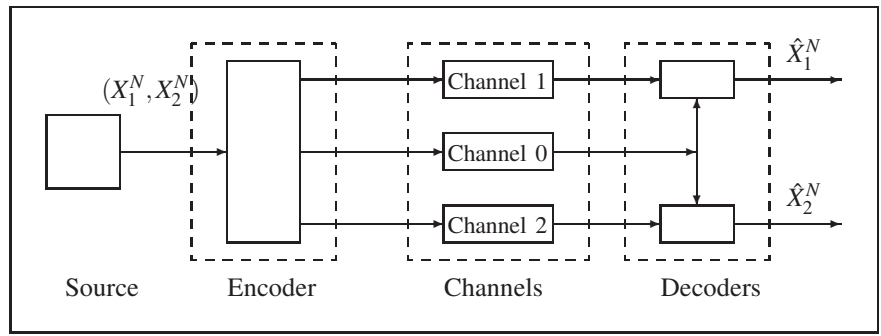
\begin{figure*}
\setlength{\unitlength}{0.24cm}
\begin{center}
\begin{picture}(40,13)(0,0)
\put(0,0){\framebox(48,18)}
\put(2,8){\framebox(4,4)}
\put(12,5){\framebox(4,10)}
\put(24,5){\framebox(6,2)}
\put(24,9){\framebox(6,2)}
\put(24,13){\framebox(6,2)}
\put(36,5){\framebox(4,2)}
\put(36,13){\framebox(4,2)}
\put(10,4){\dashbox{.5}(8,12)}
\put(22,4){\dashbox{.5}(10,12)}
\put(34,4){\dashbox{.5}(8,12)}
\put(6,10){\vector(1,0){6}}
	\put(16,6){\vector(1,0){8}}
	\put(16,10){\vector(1,0){8}}
        \put(16,14){\vector(1,0){8}}
	\put(30,6){\vector(1,0){6}}
	\put(30,10){\vector(1,0){8}}
        \put(30,14){\vector(1,0){6}}
	\put(38,10){\vector(0,-1){3}}
	\put(38,10){\vector(0,1){3}}
        \put(40,6){\vector(1,0){6}}
        \put(40,14){\vector(1,0){6}}
	\put(5,13){$(X_1^N, X_2^N)$}
	\put(43,7){$\hat{X}_2^N$}
	\put(43,15){$\hat{X}_1^N$}
	\put(24.2,5.5){{\small Channel 2}}
	\put(24.2,9.5){{\small Channel 0}}
	\put(24.2,13.5){{\small Channel 1}}
\put(3,2){Source}
\put(12,2){Encoder}
\put(24,2){Channels}
\put(36,2){Decoders}
\end{picture}
\end{center}
\caption{The Gray and Wyner source coding for a simple network \cite{gray-wyner:1974} $(X_{1,i}, X_{2,i})\sim {\bf P}_{X_1,X_2}, i=1, \ldots, N.$
}
\label{fig:gwn}
\end{figure*}

 In this  paper we show  the relevance of {\it three  fundamental concepts of statistics and probability}  to the network problems discussed above found in the report by Charalambous and van Schuppen \cite{charalambous:schuppen:2019:arxiv}  that involve a tuple of multivariate  jointly independent and identically distributed multivariate Gaussian random
variables (RVs) $(X_1^N, X_2^N)= \big\{(X_{1,i}, X_{2,i}): i=1,2, \ldots,N\big\}$,   
\begin{align}
&X_{1,i} : \Omega \rightarrow {\mathbb R}^{p_1}= {\mathbb X}_1, \ \  X_{2,i} : \Omega \rightarrow {\mathbb R}^{p_2}={\mathbb X}_2, \ \ \forall i, \label{dist_1} \\
&{\bf P}_{X_{1,i} X_{2,i}}={\bf P}_{X_1, X_2} \hso \mbox{jointly Gaussian and} \nonumber \\
&\mbox{$(X_{1,i}, X_{2,i})$ indep. of $(X_{1,j}, X_{2,j}),\hso  \forall i\neq j$}\label{dist_2}
\end{align} 
We illustrate their  application to  the calculation of rates that lie in the Gray and Wyner rate region \cite{gray-wyner:1974}  of {\it the simple network} shown in Fig.~\ref{fig:gwn}, with respect to the  average square-error distortions at the two decoders 
 \begin{align}
&{\bf E}\Big\{D_{X_i} (X_i^N, \hat{X}_i^N)\Big\}\leq \Delta_i,  \hso \Delta_i \in  [0,\infty], \hso i=1,2,\\
 & D_{X_i} (x_i^N, \hat{x}_i^N)\tri  \frac{1}{N} \sum_{j=1}^N ||x_{i,j}-\hat{x}_{i,j}||_{{\mathbb R}^{p_i}}^2, \hso i=1,2, \label{dist_3}
\end{align} 
and where $||\cdot||_{{\mathbb R}^{p_i}}^2$ are Euclidean distances on ${\mathbb R}^{p_i}, i=1,2$.\\
The  rest of this section and the remaining of the paper  is organized as follows.\\
In Section~\ref{3conc} we introduced the three concepts which are further described  in Charalambous and van Schuppen \cite{charalambous:schuppen:2019:arxiv}, in Sections~\ref{lite}-\ref{sect:wl} we  recall the Gray and Wyner characterization of the rate region \cite{gray-wyner:1974}, and the characterization of    the minimum lossy common message rate on the  Gray and Wyner rate region due to  Viswanatha, Akyol and Rose \cite{viswanatha:akyol:rose:2014}, and Xu, Liu, and Chen \cite{xu:liu:chen:2016:ieeetit}. In Section~\ref{results} we present our main results in the form of theorems.
In Section~\ref{sect:prob_st} we give  the  proofs of the main theorems, while citing  \cite{charalambous:schuppen:2019:arxiv} if necessary.

\subsection{Three Concepts of Statistics and Probability}
\label{3conc}
{\it Notation.} 
 An {\em $\rn$-valued Gaussian} RV, denoted by $X \in G(m_X, Q_X)$, 
with as parameters 
the {\em mean value} $m_X \in \rn$ and 
the {\em variance}
$Q_X \in \rnn$, $Q_X = Q_X^T \geq 0$,
is a function $X: \Omega \rightarrow \rn$ 
which is a RV and such that
the measure of this RV equals a Gaussian 
measure described by its characteristic function. 
This definition includes  $Q_X = 0$.\\
The {\it effective dimension} of the RV is denoted by 
$\dim (X) = \rank (Q_X)$. An $n\times n$ identity matrix is denoted by $I_n$.\\
A tuple of Gaussian RVs 
$(X_1, X_2)$ will be denoted this way to save space,
rather than by
\begin{align*}
  & \left(
    \begin{array}{l}
      X_1 \\ X_2
    \end{array}
    \right).
\end{align*}
Then the variance matrix of this tuple is denoted by
\begin{align*}
   & (X_1,X_2) \in G(0,Q_{(X_1,X_2)}), \\
     & Q_{(X_1,X_2)}
  =  \left(
        \begin{array}{ll}
          Q_{X_1} & Q_{X_1,X_2} \\
          Q_{X_1,X_2}^T & Q_{X_2}
        \end{array}
        \right) \in \mathbb{R}^{(p_1+p_2) \times (p_1+p_2)}.
\end{align*}
The variance $Q_{(Y_1,Y_2)}$ is distinguished 
from $Q_{Y_1,Y_2} \in \mathbb{R}^{p_1 \times p_2}$.

The {\it first concept is  Hotelling's} \cite{hotelling:1936} geometric approach to Gaussian RVs \cite{anderson:1958,gittens:1985}, where the underlying
geometric object of a Gaussian RV $Y : \Omega \rightarrow {\mathbb R}^p$ is the $\sigma-$algebra ${\cal F}^Y$ generated by $Y$. 
 A {\it basis transformation}
of such a RV is then the transformation defined by a non-singular matrix $S \in {\mathbb R}^{p\times p}$, and it then directly follows that  
${\cal F}^Y = {\cal F}^{SY}$. For the tuple of jointly Gaussian multivariate RVs $(X_1,X_2)$, a basis transformation of
this tuple consists of  a matrix composed of  two square and non-singular matrices, $(S_1, S_2)$ (see \cite[Algorithm~2.10]{charalambous:schuppen:2019:arxiv}), 
\begin{align}
 &S \tri \blockdiag ( S_1 , S_2 ), \hso X_1^c \tri S_1 X, \hso X_2^c \tri S_2 X_2, \label{meth_1a_i} \\
&{\cal F}^{X_1} = {\cal F}^{S_1 X_1}, \ \ {\cal F}^{X_2} = {\cal F}^{S_2 X_2 }.\label{meth_1c_i}
\end{align}
$S$   maps $(X_1, X_2)$ into  the so-called canonical form of the tuple of RVs  (the full specification is given in  \cite[Section~2.2, Definition~2.2]{charalambous:schuppen:2019:arxiv}), which identifies identical, correlated, and private information, as interpreted  in the table below,
 \begin{center}
\begin{tabular}{|l|l|}
\hline 
$X_{11}^c = X_{21}^c-\as$ & {\em identical information}  of $X_1^c$ and $X_2^c$ \\
$X_{12}^c$          & {\em correlated information} of $X_1^c$ w.r.t $X_2^c$ \\
$X_{13}^c$          & {\em private information}    of $X_1^c$ w.r.t $X_2^c$ \\ \hline
$X_{21}^c = X_{11}^c-\as$ & {\em identical information}  of $X_1^c$ and $X_2^c$ \\
$X_{22}^c$          & {\em correlated information} of $X_2^c$ w.r.t  $X_1^c$ \\
$X_{23}^c$          & {\em private information}    of $X_2^c$ w.r.t $X_1^c$ \\ \hline
\end{tabular}
\end{center}
\par\vspace{1\baselineskip}\par\noindent
 where 
  \begin{align}
& X_{ij}^c: \Omega \rightarrow \mathbb{R}^{p_{ij}}, ~
        i =1, 2, ~ j = 1, 2, 3, \\
&p_{11}=p_{21}, \hso  p_{12}=p_{22}=n,\\
&p_1=p_{11}+p_{12}+p_{13}, \hso p_2=p_{21}+p_{22}+p_{23},  \label{cvf_3_a_i}\\
& S_1X_1=(X_{11}^c,X_{12}^c, X_{13}^c),  \hso S_2X_2=(X_{21}^c,X_{22}^c,X_{23}^c), \label{cvf_1_a_i}\\
 &X_{11}^c=X_{21}^c-\as, \hso  X_{11}^c, X_{21}^c \in G(0,I_{p_{11}}),\\
&\mbox{$X_{13}^c \in G(0,I_{p_{13}})$ and $X_{23}^c \in G(0,I_{p_{23}})$ are independent} \label{cvf_2_i}  \\
&\mbox{$X_{12}^c \in G(0,I_{p_{12}})$ and $X_{22}^c \in G(0,I_{p_{22}})$ are correlated}, \label{cvf_2a_i}  \\
&{\bf E}[X_{12}^c (X_{22}^c)^T]=  D=\diag (d_1,\ldots, d_{p_{12}}), \: d_i \in (0,1)\; \forall i. \label{cvf_3_i}
 \end{align}
The entries of $D$ are called {\em the canonical correlation coefficients}. 
For $X_{11}^c = X_{21}^c-\as$ the term {\em identical information} is used. The linear transformation $S = \blockdiag ( S_1 , S_2 )$ is equivalent to a pre-processing
of $(X_1,X_2)$ by a 
linear pre-encoder (see \cite{charalambous:schuppen:2019:arxiv} for  applications to network problems).

The expression of mutual information between $X_1$ and $X_2$, denoted by  $I(X_1;X_2)$,  as a function of  the canonical correlation coefficients, discussed in \cite{gelfand:yaglom:1959} is  given in Theorem~\ref{them_mi}.

The {\it second concept is van Putten's and van Schuppen's} \cite{putten:schuppen:1985}  
 parametrization  of the family of all jointly Gaussian probability distributions ${\bf P}_{X_1,X_2, W}$ by an auxiliary Gaussian  RV $W: \Omega \rar {\mathbb R}^k= {\mathbb W}$ that makes $X_1$ and $X_2$ conditional independent, defined by  
\begin{align}
&{\cal P}^{CIG} \tri \Big\{ {\bf P}_{X_1, X_2, W}\Big| \ \ {\bf P}_{X_1, X_2|W}={\bf P}_{X_1|W} {\bf P}_{X_2|W}, \nonumber \\
& \mbox{the ${\mathbb X}_1\times {\mathbb X}_2-$marginal dist. of     ${\bf P}_{X_1, X_2, W}$ is the fixed dist.} \nonumber \\
&\mbox{ ${\bf P}_{X_1, X_2}$, and   $(X_1, X_2,W)$  is jointly Gaussian}   \Big\}
\end{align}
and its  subset ${\cal P}_{min}^{CIG}$ of the set ${\cal P}^{CIG}$, with the additional constraint  that the dimension of the RV $W$
is minimal while all other conditions hold.  The parametrizaion
is in terms of a set of matrices.  Consequences  are  found in \cite[Section~2.3]{charalambous:schuppen:2019:arxiv}.

The {\it third concept is the weak stochastic realization}  of RVs $(X_1, X_2,W)$ that induces distributions ${\bf P}_{X_1, X_2, W}$ in the sets ${\cal P}^{CIG}$ and ${\cal P}_{min}^{CIG}$ (see \cite[Def. 2.17 and Prop. 2.18]{putten:schuppen:1985} and \cite[Def.~2.17 and Prop.~2.18]{charalambous:schuppen:2019:arxiv}).

Theorem~\ref{cor:commoninfogrvcorrelated} (our main theorem)  gives as a special case (part (d))  an achievable lower bound 
on  Wyner's single letter {\it information theoretic characterization of  common information}:  
\begin{align}
C_W(X_1, X_2)\tri \inf_{{\bf P}_{X_1, X_2, W}:\: {\bf P}_{X_1, X_2|W}={\bf P}_{X_1|W} {\bf P}_{X_2|W}} I(X_1, X_2; W) \label{eq_41_in}
\end{align}
and the weak stochastic realization of  RVs $(X_1, X_2,W)$ that induce distributions ${\bf P}_{X_1, X_2, W}$ in the sets ${\cal P}^{CIG}$ and ${\cal P}_{min}^{CIG}$. 

\subsection{The Gray and Wyner Lossy Rate Region}
\label{lite}
Now, we describe our results with respect to the  fundamental question posed by Gray and Wyner \cite{gray-wyner:1974} for the simple network shown in Fig.~\ref{fig:gwn}, which  is:  determine which channel capacitity triples 
$(C_0, C_1, C_2)$ are necessary and sufficient for each sequence $(X_1^N, X_2^N)$ to be reliably reproduced at the intended decoders, while satisfying the average distortions with respect to single letter distortion functions $D_{X_i} (x_i^N, \hat{x}_i^N)\tri \frac{1}{N} \sum_{t=1}^n d_{X_i}(x_{i,t} \hat{x}_{i,t}), i=1,2$. Gray and Wyner characterized the {\it operational rate region}, denoted by  ${\cal R}_{GW}(\Delta_1, \Delta_2)$   by a coding scheme that uses the auxiliary RV $W: \Omega \rar {\mathbb W}$, as described below. Define the family of probability distributions
\begin{align*}
{\cal P} \triangleq 
    & \Big\{ 
        \begin{array}{l}
          {\bf P}_{X_1, X_2, W}, \ \ 
          x_1 \in {\mathbb X}_1, ~ x_2 \in {\mathbb X}_2, ~ w \in {\mathbb W} ~
          \Big| \\
          {\bf P}_{X_1, X_2, W}(x_1,x_2,\infty) = {\bf P}_{X_1, X_2}
        \end{array}
      \Big\}
\end{align*}
for some auxiliary random variable  $W$.

Theorem 8 in \cite{gray-wyner:1974}: 
Let ${\cal R}_{GW}(\Delta_1, \Delta_2)$ denote the Gray and Wyner rate region. 
 Suppose there exists $\hat{x}_i \in \hat{\mathbb X}_i$ such that ${\bf E}\{d_{X_i}(X_i, \hat{x}_i)\}< \infty$,   $i=1,2$. For each ${\bf P}_{X_1, X_2, W} \in {\cal P}$ and $\Delta_1 \geq 0, \Delta_2 \geq 0$, define the subset of Euclidean $3-$D space 
 \begin{align}
  {\cal R}_{GW}^{{\bf P}_{X_1,X_2,W}}&(\Delta_1, \Delta_2) = \Big\{\Big(R_0,R_1,R_2\Big): \ \       R_0 \geq I(X_1, X_2; W), \nonumber \\ 
& R_1 \geq R_{X_1|W}(\Delta_1), \ \ R_2 \geq R_{X_2|W}(\Delta_2) \Big\} \label{eq_32} 
\end{align} 
where $R_{X_i|W}(\Delta_i)$ is rate distortion function (RDF) of $X_i$, conditioned on $W$, at decoder $i$,  $i=1,2$,  and $R_{X_1,X_2}(\Delta_1,\Delta_2)$ is the joint RDF of joint decoding of $(X_1, X_2)$. 
Let
\begin{align}
{\cal R}_{GW}^{*}(\Delta_1, \Delta_2) \tri \Big(\bigcup_{ {\bf P}_{X_1,X_2, W} \in {\cal P}} {\cal R}_{GW}^{{\bf P}_{X_1,X_2,W}}(\Delta_1, \Delta_2)\Big)^c
\end{align}
where $\big(\cdot\big)^c$ denotes the closure of the  indicated set. 
Then the achievable Gray-Wyner lossy rate region  is given by
\begin{align}
{\cal R}_{GW}(\Delta_1, \Delta_2)={\cal R}_{GW}^{*}(\Delta_1, \Delta_2).  \label{eq_31}
\end{align}

By  \cite[Theorem~6]{gray-wyner:1974} if   $(R_0, R_1, R_2) \in {\cal R}_{GW}(\Delta_1, \Delta_2)$, then
\begin{align}
& R_0 + R_1 +R_2 \geq R_{X_1, X_2}(\Delta_1, \Delta_2),   \label{eq_32a}  \\
& R_0 +R_1  \geq R_{X_1}(\Delta_1), \hso  R_0 +R_2  \geq R_{X_2}(\Delta_2) \label{eq_32c} 
\end{align}
(\ref{eq_32a}) is called the {\em Pangloss Bound} 
 of ${\cal R}_{GW}(\Delta_1, \Delta_2)$, and the set of triples $(R_0, R_1, R_2) \in {\cal R}_{GW}(\Delta_1, \Delta_2)$ that satisfy  $R_0+R_1+R_2= R_{X_1, X_2}(\Delta_1, \Delta_2)$ 
 the {\em Pangloss Plane}.

Theorem~\ref{cor:commoninfogrvcorrelated} is our main  theorem for set up (\ref{dist_1})-(\ref{dist_3}). From this theorem follows Proposition~\ref{prop_1} that parametrizes the region ${\cal R}_{GW}(\Delta_1, \Delta_2)$ by a Gaussian RV $W$, and the weak stochastic realization of the joint  distribution of $(X_1, X_2, W)$.

\subsection{Wyner's Lossy Common Information} 
\label{sect:wl}
 Viswanatha, Akyol, and Rose \cite{viswanatha:akyol:rose:2014}, and Xu, Liu, and Chen \cite{xu:liu:chen:2016:ieeetit}, characterized   the minimum lossy common message rate on the   rate region ${\cal R}_{GW}(\Delta_1, \Delta_2)$, as follows.
 
Theorem 4 in \cite{xu:liu:chen:2016:ieeetit}: Let $C_{GW}(X_1, X_2; \Delta_1, \Delta_2)$ denote the minimum common message rate $R_0$ on the Gray and Wyner lossy rate region  ${\cal R}_{GW}(\Delta_1,\Delta_2)$, with sum rate not exceeding the joint rate distortion function $R_{X_1,X_2}(\Delta_1, \Delta_2)$.   \\
Then  $C_{GW}(X_1, X_2; \Delta_1, \Delta_2)$  is characterized by  
\begin{align}
C_{GW}(X_1, X_2; \Delta_1, \Delta_2) \tri \inf\: I(X_1, X_2; W)
\end{align}
such that the following identity holds
\bea
R_{X_1|W}(\Delta_1)+R_{X_2|W}(\Delta_2)+ I(X_1, X_2; W)=R_{X_1, X_2}(\Delta_1, \Delta_2) \label{equality_1}
\eea
where the infimum is over all RVs $W$ in ${\mathbb W}$, which parametrize the source distribution via ${\bf P}_{X_1,X_2,W}$, having a ${\mathbb X}_1\times {\mathbb X}_2-$marginal  source
distribution ${\bf P}_{X_1,X_2}$, and induce joint distributions ${\bf P}_{W,X_1,X_2,\hat{X}_1,\hat{X}_2} $ which satisfy the constraint.

$C_{GW}(X_1, X_2; \Delta_1, \Delta_2)$ is  also given the interpretation of   Wyner's lossy common information, due to its operational meaning \cite{viswanatha:akyol:rose:2014,xu:liu:chen:2016:ieeetit}.  We should mention that from   Appendix B in \cite{xu:liu:chen:2016:ieeetit} it follows that a  necessary condition for the equality constraint (\ref{equality_1}) is 
$R_{X_1, X_2|W}(\Delta_1, \Delta_2)=R_{X_1|W}(\Delta_1)+ R_{X_2|W}(\Delta_2)$, and  sufficient condition for this equality  to hold is the conditional independence condition \cite{xu:liu:chen:2016:ieeetit}: ${\bf P}_{X_1,X_2|W}={\bf P}_{X_1|W} {\bf P}_{X_2|W}$.  Hence, a sufficient condition for  any rate $(R_0,R_1,R_2) \in  {\cal R}_{GW}(\Delta_1,\Delta_2)$ to lie on the Pangloss plane, i.e., to satisfy (\ref{equality_1})  is the  conditional independence.

It is shown in \cite{viswanatha:akyol:rose:2014,xu:liu:chen:2016:ieeetit}, that there exists a distortion region ${\cal  D}_W \subseteq [0,\infty]\times [0,\infty]$ such that $C_{GW}(X_1, X_2; \Delta_1, \Delta_2) = C_W(X_1, X_2)$, i.e., it is independent of the distortions $(\Delta_1, \Delta_2)$,  i.e. it equals  the Wyner's  information theoretic characterization of  common information  defined by (\ref{eq_41_in}).  

From Theorem~\ref{cor:commoninfogrvcorrelated} follows Theorem~\ref{thm:r0} that gives the closed form expression of $C_{GW}(X_1, X_2; \Delta_1, \Delta_2)= C_W(X_1, X_2)$ and identifies  the region ${\cal  D}_W$, for the multivariate Gaussian RVs $(X_1, X_2)$ with respect to the avarage  distortions  (\ref{dist_1})-(\ref{dist_3}).

\section{Main Results}
\label{results}
Given the tuple of multivariate Gaussian RVs and distortion functions (\ref{dist_1})-(\ref{dist_3}),   the main contributions of the paper are:\\
(1) the theorem and the proof of Wyner's common information (information definition).
The existing proof of this result in \cite{satpathy:cuff:2012}  is incomplete (see discussion below Theorem~\ref{cor:commoninfogrvcorrelated}).

(2) Paremetrization of rate triples $(R_0, R_1, R_2) \in {\cal R}_{GW}(\Delta_1, \Delta_2)$,  and Wyner's lossy common  information.

Below we state  the expression of mutual information as a function of  the canonical correlation coefficients, discussed in Gelfand and Yaglom \cite{gelfand:yaglom:1959}.

\begin{theorem}\label{them_mi}
Consider a tuple of multivariable jointly Gaussian RVs
$X_1: \Omega \rar {\mathbb R}^{p_1}, X_2 :\Omega \rar {\mathbb R}^{p_2}$, $(X_1,X_2) \in G(0,Q_{(X_1,X_2)})$. 
Compute the canonical variable form of the tuple
of Gaussian RVs
according to Algorithm~2.2 of \cite{charalambous:schuppen:2019:arxiv}.
This yields the indices
$p_{11} = p_{21}$, $p_{12} = p_{22}$, $p_{13}$, $p_{23}$, and 
$n = p_{11} + p_{12} = p_{21} + p_{22}$
and the diagonal matrix $D$ with canonical correlation coefficients 
$d_i \in (0,1)$ for $i = 1, \ldots, n$ (as in \cite[Definition~2.2]{charalambous:schuppen:2019:arxiv}).\\
Then mutual information  $I(X_1; X_2)$ is given by  the formula,
\begin{align*}
      I(X_1;X_2)
   = & \left\{
        \begin{array}{ll}
            0, 
           & 
              0 = p_{11}=p_{12}, \\
            -\frac{1}{2} \sum_{i=1}^n 
            \ln
            \left(1-d_i^2
            \right), 
          & 
              0 = p_{11}, 
              p_{12}  > 0,  \\
            \infty,  
          & 
              p_{11}  > 0
        \end{array}
        \right.
\end{align*}
where $d_i$ are the canonical correlation coefficients.
\end{theorem}

$I(X_1; X_2)$ is a generalization of the well-known formula of a tuple of scalar RVs, i.e., $p_1=p_2=1$, $I(X_1;X_2)=-\frac{1}{2} 
            \ln
            \left(1-\rho^2
            \right)$, where $\rho \tri {\bf E}\big\{X_1 X_2\} \in [-1,1]$ is the correlation coefficient.

The case $p_{11}=p_{21}>0$ gives $I(X_1; X_2)=+\infty$; if such components are present they should be removed.  Hence, we  state the next theorem under the restriction $p_{11}=p_{21}=0$.

\begin{theorem}\label{cor:commoninfogrvcorrelated}
Consider a tuple of multivariable jointly Gaussian RVs
$X_1: \Omega \rar {\mathbb R}^{p_1}, X_2 :\Omega \rar {\mathbb R}^{p_2}$, $(X_1,X_2) \in G(0,Q_{(X_1,X_2)})$ and without loss of generality assume $S = \blockdiag ( S_1 , S_2 )$ produces  a canonical variable form  such that  $p_{11}=p_{21}=0$ (see \cite[Definition~2.2]{charalambous:schuppen:2019:arxiv}). \\
For any joint distrubution ${\bf P}_{X_1,X_2,W}$ parametrized by  an arbitrary RV $W: \Omega \rar {\mathbb R}^k$ with fixed marginal distribution ${\bf P}_{X_1,X_2}=G(0,Q_{(X_1,X_2)})$ the following hold.\\
(a) The mutual information  $I(X_1,X_2; W)$ satisfies
\begin{align}
 I(&X_1,X_2;W)=I(X_{12}^c,X_{22}^c; W), \hso p_{12}=p_{22}=n  \label{thm_1_eq_1}  \\
   \geq & H(X_{12}^c,X_{22}^c) -  H(X_{12}^c|W) - H(X_{22}^c|W), \label{ine_1} \\
   =  & \frac{1}{2} \sum_{i=1}^n \ln (1-d_i^2) 
        \nonumber \\
        &- \frac{1}{2} \ln (\det ( [I - D^{1/2} Q_W^{-1} D^{1/2} ]
                                  [ I - D^{1/2} Q_W D^{1/2} ] 
                                )
                          )\label{thm_1_eq_2}
\end{align}
where the lower bound  is parametrized by $Q_W \in {\bf Q_W}$, 
\bea
{\bf Q_W}
   =  \Big\{ Q_W \in \rnn |\; Q_W = Q_W^T, ~
                          0 < D \leq Q_W \leq D^{-1} 
        \big\} \label{eq:pci_1_in} 
\eea
and such that ${\bf P}_{X_{11}^c,X_{22}^c, W}$ is jointly Gaussian.\\
(b) The lower bound in (\ref{ine_1}) is achieved if ${\bf P}_{X_1,X_2, W}$ is jointly Gaussian and ${\bf P}_{X_{12}^c,X_{22}^c| W}={\bf P}_{X_{12}^c| W}{\bf P}_{X_{22}^c| W}, W : \Omega\rar  {\mathbb R}^n$,  and a realization of the RVs $(X_{12}^c, X_{22}^c)$  which achieves the lower bound is 
\begin{align}
      X_{12}^c 
   = & Q_{X_{12}^c, W}Q_W^{-1}W+Z_1, \label{rep_g_s1_in_a}\\
      Q_{X_{12}^c, W}
   = & D^{1/2}, \ \   Z_1 \in G(0,(I-D^{1/2}Q_W^{-1}D^{1/2}  )), 
          \label{rep_g_s1_in}  \\
      X_{22}^c 
   = & Q_{X_{22}^c,W} Q_W^{-1} W + Z_2 \label{rep_g_s2_in}\\
      Q_{X_{22}^c, W}
   = & D^{1/2}Q_W, \ \   Z_2 \in G(0,(I-D^{1/2}Q_WD^{1/2}  )),   \\
     & (Z_1,Z_2,W), ~ \mbox{are independent.} \label{rep_g_s3} 
\end{align}  
(c) A lower bound on (\ref{thm_1_eq_2}) occurs if $Q_W=Q_{W^*} \in {\bf Q}_W$ is diagonal, i.e., $Q_{W^*}=\diag (Q_{W_1^*},\ldots, Q_{W_n^*}), \: d_i \leq Q_{W_i^*} \leq d_i^{-1}, \forall i$, and it is achieved by realization (\ref{rep_g_s1_in_a})-(\ref{rep_g_s3}), with $Q_W=Q_{W^*}$. \\
(d) Wyner's information common information  is given by 
\begin{align}
     C_W(X_1,X_2)
   = \left\{ \begin{array}{cc} \frac{1}{2} \sum_{i=1}^n 
        \ln
        \left(
        \frac{1+d_i}{1-d_i}
        \right) 
         \in (0,\infty) &  \mbox{if $n>0$}  \\
         0 & \mbox{if $n=0$}
\end{array} \right.         
          \label{lci_1_in}
\end{align}
and it is achieved by a Gaussian RV $W=W^* \in G(0,Q_{W^*}), W^*: \Omega \rar {\mathbb R}^n$, $Q_{W^*}=I_n$ an $n\times n$ identity covariance matrix, and the realization of part (b) with $Q_W=I_n$.
\end{theorem}

The characterization of the subset ${\cal P}_{min}^{CIG}$ of the set ${\cal P}^{CIG}$ of two RVs $(X_1, X_2)$ in canonical variable form by  the set ${\bf Q_W}$ is due to Van Putten and Van Schuppen \cite{putten:schuppen:1985}. 

In \cite{satpathy:cuff:2012} the proof of (\ref{lci_1_in}) is incomplete
because there is no optimization over the set of measures $Q_W \in {\bf Q_W}$
achieving the conditional independence. 
In that reference there is an assumption
that three cross-covariances can be simultaneously diagonalized.
which is not true in general. 
This assumption implies that case (d) of the above theorem holds.
This assumption is repeated in \cite{veld-gastpar:2016}.

From Theorem~\ref{cor:commoninfogrvcorrelated} follows directly the proposition  below.

\begin{proposition}
\label{prop_1}
 Consider the statement of Theorem~\ref{cor:commoninfogrvcorrelated}, with $(X_1, X_2)$ in canonical variable form. Then ${\cal R}_{GW}(\Delta_1, \Delta_2)$
is determined from 
\begin{align*}
T(\alpha_1, \alpha_2) =& \inf_{ {\bf Q_W}} \Big\{I(X_1, X_2; W)+\alpha_1 R_{X_1|W}(d_1) + \alpha_2 R_{X_2|W}(d_2)\Big\}
\end{align*}
$0\leq \alpha_i\leq 1, i=1,2,  \alpha_1+\alpha_2\geq 1$, 
and the infimum occurs at the  diagonal $Q_W=Q_{W^*} \in {\bf Q_W}$ of Theorem~\ref{cor:commoninfogrvcorrelated}, part (c). Moreover, $R_{X_i|W}(d_i), i=1,2$ is given by 
\begin{align}
R_{X_i|W}(\Delta_i)=&  \inf_{ \sum_{j=1}^n \Delta_{i,j}=\Delta_i}  \frac{1}{2} \sum_{j=1}^n  \log \Big( \frac{(1-d_{j}/Q_{W_j}^*)}{\Delta_{i,j} }\Big)^+  \label{jrdf_1}
\end{align}
where $(\cdot)^+ \tri \max\{1, \cdot\}$,  $
{\bf E}||X_{i2}^c- \hat{X}_{i2}^c||_{{\mathbb R}^{n}}^2=\sum_{j=1}^n \Delta_{i,j}=\Delta_i, i=1,2$, and  the water-filling equations hold:
\begin{align}
&
\Delta_{i,j} = \left\{ \begin{array}{cc} \lambda, & \lambda < 1-d_{j}  \  \\
1-d_{j}, & \lambda \geq 1-d_{j}, \end{array} \right. \: \Delta_i\in (0,\infty), \: i=1,2.
\end{align}
\end{proposition}
{\bf Proof}  Follows from Gray and Wyner \cite[(4) of page 1703,  eqn(42)]{gray-wyner:1974} and Theorem~\ref{cor:commoninfogrvcorrelated}. (\ref{jrdf_1}) follows from RDF of Gaussian RVs.

\begin{theorem}\label{thm:r0}
Consider the  tuple of jointly Gaussian RVs of Theorem~\ref{cor:commoninfogrvcorrelated}.  
Then  
\begin{align}
C_{GW}(X_1,&X_2;\Delta_1,\Delta_2)= C_W(X_1,X_2)\\
   =&  \frac{1}{2} \sum_{j=1}^n 
        \ln
        \left(
        \frac{1+d_j}{1-d_j}
        \right), \ \   (\Delta_1, \Delta_2)\in {\cal D}_W  \label{wlc}\\
{\cal D}_W \tri & \Big\{(\Delta_1, \Delta_2)\in [0,\infty]\times [0,\infty]\Big| \; 0\leq \Delta_1 \leq \sum_{j=1}^n(1-d_j), \nonumber  \\
 &0\leq \Delta_2 \leq \sum_{j=1}^n(1-d_j)\Big\}, \; {d_j \in (0,1)}, j=1, \ldots, n.\nonumber 
\end{align} 
\end{theorem}

Formula (\ref{wlc}) is a generalization of the analogous formula derived in \cite{gray-wyner:1974,viswanatha:akyol:rose:2014,xu:liu:chen:2016:ieeetit}, for a  tuple of  jointly Gaussian scalar  RVs $(X_1, X_2)$, zero mean, ${\bf E}[X_1^2]={\bf E}[X_2^2]=1$, ${\bf E}[X_1 X_2]=\rho \in [0,1]$.

\section{Proofs of main Theorems}
\label{sect:prob_st}
We present in this section additional exposition on the Concepts of Section~\ref{3conc}, and  outlines of the proofs of the main theorems (see \cite{charalambous:schuppen:2019:arxiv} for additional exposition).

\subsection{Further Discussion on the Three Conecpts}
First we state  a few facts. \\
(A1) The parametrization of  the family of Gaussian probability distributions
${\cal P}^{CIG}$ and ${\cal P}_{min}^{CIG}$ require the solution of the weak stochastic realization problem of Gaussian RVs (defined by Problem~2.15 in \cite{charalambous:schuppen:2019:arxiv}) given in  \cite[Theorem~4.2]{putten:schuppen:1983} (see also \cite[Theorem~3.8]{charalambous:schuppen:2019:arxiv}), and reproduced below.  

\begin{theorem}\cite[Theorem 4.2]{putten:schuppen:1983}
\label{them_putten:schuppen:1983}
Consider a tuple $(X_1,X_2)$ of Gaussian RVs
in the canonical variable form.
Restrict attention to the correlated parts of these RVs, as follows:
\begin{align}
    & (X_1,X_2) \in G(0,Q_{(X_1,X_2)}) = {\bf P}_0, \ \
         X_1,X_2: \Omega \rightarrow \rn,   \label{ci_par_1}   \\
     & Q_{(x_1,x_2)}
   = \left(
        \begin{array}{ll}
          I & D \\
          D & I
        \end{array}
        \right), \; p_{11}=p_{21}=0, p_{13}=p_{23}=0,  \label{ci_par_2}  \\
      &D
 = \diag (d_1,  \ldots, d_n) \in \rnn, ~
        1 > d_1\geq \ldots \geq d_n > 0.
\end{align}
\begin{itemize}
\item[(a)] There exists a probability measure ${\bf P}_1$, and  a triple of Gaussian RVs
$X_1,X_2,~~ W: \Omega \rightarrow \rn$ defined on it, such that (i) 
${\bf P}_1|_{(X_1,X_2)} = {\bf P}_0$ and (ii)
${X_1}$ and $X_2$ are conditional independent given $W$ with  $W$ having  minimal dimension.  

\item[(b)] There exist a family of Gaussian measures denoted by ${\bf P_{ci}}\subseteq {\cal P}_{min}^{CIG}$, that satisfy (i) and (ii) of (a), and moreover this family is  parametrized by the matrices and sets:
\begin{align}
   & G(0,Q_s(Q_W)), ~
       Q_W \in {\bf Q_W},  \label{eq:gtriple} \\
    & 
     Q_s(Q_W) =
        \left(
        \begin{array}{lll}
          I       & D           & D^{1/2} \\
          D       & I           & D^{1/2} Q_W \\
          D^{1/2} & Q_W D^{1/2} & Q_W 
        \end{array}
        \right),\label{eq:gtriple_1} \\
     & {\bf Q_W}
   =  \Big\{ Q_W \in \rnn \Big|\; Q_W = Q_W^T, ~
                          0 < D \leq Q_W \leq D^{-1} 
        \big\}, \label{eq:pci_1} \\
     &   {\bf P_{ci}}
  =  \Big\{ G(0,Q_s(Q_W)) ~ \mbox{on} ~  
           (\mathbb{R}^{3n}, {\cal B}(\mathbb{R}^{3n})) \Big|\;  Q_W \in {\bf Q_W} 
        \Big\} \nonumber 
\end{align}
and ${\bf P_{ci}} \subseteq {\cal P}_{min}^{CIG}$. 
\end{itemize}
\end{theorem}

%
%

(A2) The weak stochastic realization of a Gaussian measure $G(0,Q_0)$ on the Borel space
$(\mathbb{R}^{p_1+p_2}, {\cal B}(\mathbb{R}^{p_1+p_2}))$ is then defined and characterized as in Def.~2.17 and Prop.~2.18, Alg.~3.4 of \cite{charalambous:schuppen:2019:arxiv}.

\subsection{Proofs of Main Theorems}
(B)  For  the calculatation of $ C_W(X_1, X_2)$ via Theorem~\ref{cor:commoninfogrvcorrelated} and 
$C_{GW}(X_1,X_2;\Delta_1,\Delta_2)$ via Theorem~\ref{thm:r0} 
 it is sufficient
to impose the conditional independence   ${\bf P}_{X_1, X_2|W}={\bf P}_{X_1|W} {\bf P}_{X_2|W}$, due to,  \\
(a) the well-known inequality 
\begin{align}
I(X_1,X_2;W) 
\geq   H(X_1,X_2)-H(X_1|W)-H(X_2|W)
\end{align}
which is achieved if ${\bf P}_{X_1, X_2|W}={\bf P}_{X_1|W} {\bf P}_{X_2|W}$. 

(b) A  necessary condition for the equality constraint (\ref{equality_1}) to hold   (see Appendix B in \cite{xu:liu:chen:2016:ieeetit}) is 
\begin{align}
R_{X_1, X_2|W}(\Delta_1, \Delta_2)=R_{X_1|W}(\Delta_1)+ R_{X_2|W}(\Delta_2). \label{nec_equality_1}
\end{align}
Further, a sufficient condition for (\ref{nec_equality_1}) to hold is the conditional independence condition \cite{xu:liu:chen:2016:ieeetit}: ${\bf P}_{X_1,X_2|W}={\bf P}_{X_1|W} {\bf P}_{X_2|W}$. 
 a sufficient condition for  any rate $(R_0,R_1,R_2) \in  {\cal R}_{GW}(\Delta_1,\Delta_2)$ to lie on the Pangloss plane is the  conditional independence. 

(c) For jointly Gaussian RVs $(X_1,X_2)$ with square-error distortion, then by the maximum
entropy principle the optimal joint distribution ${\bf P}_{X_1, X_2,\hat{X}_1,\hat{X}_2,W}$ of the optimization problem $C_{GW}(X_1,X_2;\Delta_1,\Delta_2)$ is  Gaussian.

(d) The characterization of Wyner's information  common information $C_W(X_1, X_2)$ for jointly Gaussian multivariate RVs $(X_1, X_2)$ occurs in the set of jointly Gaussian RVs $(X_1, X_2, W)$ such that ${\bf P}_{X_1,X_2|W}={\bf P}_{X_1|W} {\bf P}_{X_2|W}  $ and $C_W(X_1, X_2)$ is invariant with respect to Hotelling's  nonsingular basis transformation   

(e) For data (\ref{dist_1})-(\ref{dist_3}), any rate triple $(R_0,R_1, R_2)$ that belongs to ${\cal R}_{GW}(\Delta_1, \Delta_2)$, characterized by (\ref{eq_32})-(\ref{eq_32c}),  is equivalently computed by transforming the tuple $(X_1,X_2)$
into  their canonical variable form  (\ref{meth_1a_i})-(\ref{cvf_3_i}).

\begin{remark} Theorem~\ref{them_putten:schuppen:1983} is a parametrization of  the familiy of Gaussian measures ${\bf P_{ci}}\subseteq {\cal P}_{min}^{CIG}$ by $Q_W$ and $Q_{W^*}$. \\
(a) Theorem~\ref{them_putten:schuppen:1983}  applies to other network problems, i.e.,   the Gaussian   MACs  by incorporating average power constraints.\\
(b) The weak stochastic realization of RVs  $(X_1, X_2)$, in terms of the random variable $W$ is given in Theorem~\ref{cor:commoninfogrvcorrelated}. \\
(c) An alternative proof of Proposition~\ref{prop_1}, i.e.,  that ${\cal R}_{GW}(\Delta_1, \Delta_2)={\cal R}_{GW}^{*}(\Delta_1, \Delta_2)$ is generated from distributions  ${\bf P_{ci}}\subseteq {\cal P}_{min}^{CIG}\subseteq {\cal P}$ is given in  \cite{cdc:jhv:2019}. 
\end{remark}

%
%
%
%


{\bf Proof of Theorem~\ref{them_mi}:} Follows from the non-singular transformations (\ref{meth_1a_i}), (\ref{meth_1c_i}), and chain rule of mutual information applied to $I(X_1; X_2)=I(S_1 X_1; S_1X_2)=I(X_{11}^c, X_{12}^c, X_{13}^c;X_{21}^c, X_{22}^c, X_{23}^c)$.  

{\bf Proof of Theorem~\ref{cor:commoninfogrvcorrelated}:} An alternative derivation  based on inequalities of linear algebra is given in Theorem~3.11 of \cite{charalambous:schuppen:2019:arxiv}, and is based on
\cite[Theorem~9.E.6]{marshall:olkin:1979}, with reference to Hua LooKeng \cite{hua:1955a}. Below, we present a simplified derivation. \\
(a) Equality  (\ref{thm_1_eq_1}) follows from  the non-singular transformations (\ref{meth_1a_i}), (\ref{meth_1c_i}); 
inequality (\ref{ine_1}) is due to (B).(a); (\ref{thm_1_eq_2}) follows by evaluation of entropies;  (\ref{eq:pci_1_in}) is due to Theorem~\ref{them_putten:schuppen:1983}. (b) The lower bound is achieved by the maximum entropy principle of Gaussian RVs, and the realization is due to Theorem~\ref{them_putten:schuppen:1983}. (c) We identify a further lower bound  on  the  second right-hand-side term of  (\ref{thm_1_eq_2})    that depends on $Q_W \in {\bf Q_W}$ (and corresponds to $-H(X_{12}^c|W)-H(X_{22}^c|W)$, by letting $Q_W=Q_{W^*}$.  By the chain rule of entropy then 
\begin{align*}
H(X_{12}^c|W)
=& \sum_{j=1}^n H(X_{12,j}^c|W_1, \ldots, W_n,X_{12,1}^c, X_{12,2}^c, \ldots, X_{12,j-1}^c )\\
\leq & \sum_{j=1}^n H(X_{12,j}^c| W_j )
\end{align*}
and the upper bound is achieved if $(X_{12,j}, W_j), j=1, \ldots, n$ are jointly independent, hence $Q_W=Q_{W^*}$. Similarly, the upper bound $H(X_{22}^c|W)\leq \sum_{j=1}^n H(X_{22,j}^c| W_j )$ is achieved if  $(X_{22,j}, W_j), j=1, \ldots, n$ are jointly independent, i.e., $Q_W=Q_{W^*}$. Such joint distribution is induced by the  realization of  part (b),   with $Q_W=Q_{W^*}$. (d) By part (c), and  simple algebra we can show that the optimal $Q_{W^*}$ for $C_W(X_1, X_2)$ is $Q_{W^*}=I_n \in {\bf Q_{W^*}}$  and then follows (\ref{lci_1_in}). 

{\bf Proof of Theorem~\ref{thm:r0}:}
 A direct  way to prove the statement is to compute the  rate distortion functions $R_{X_i}(\Delta_i), R_{X_i|W}(\Delta_i), i=1,2$ and $R_{X_1, X_2}(\Delta_1, \Delta_2)$, using   the weak stochastic realization of Theorem~\ref{them_putten:schuppen:1983}.(b), and then verify that identity (\ref{equality_1}) holds, i.e., $R_{X_1|W}(\Delta_1)+R_{X_2|W}(\Delta_2)+ I(X_1, X_2; W)=R_{X_1, X_2}(\Delta_1, \Delta_2)$ for $(\Delta_1,\Delta_2) \in {\cal D}_W$, for the choice $W=W^*$ with $Q_{W^*}=I_n$ given in  Theorem~\ref{cor:commoninfogrvcorrelated}.(b). 


\section{Concluding Remarks}\label{sec:concludingremarks}
This paper calculates rates on the Gray and Wyner lossy rate region of a tuple of jointly Gaussian RVs, $X_1 : \Omega \rar {\mathbb R}^{p_1}, X_2 : \Omega \rar {\mathbb R}^{p_2}$ with square-error fidelity at the two decoders, by making use of van Putten's and van Schuppen's  \cite{putten:schuppen:1985} parametrization of all jointly Gaussian  distributions ${\bf P}_{X_1, X_2,W}$,   by another Gaussian RV $W : \Omega \rar {\mathbb R}^{n}$,   such that ${\bf P}_{X_1, X_2|W}= {\bf P}_{X_1|W}{\bf P}_{X_2|W}$,   and their weak stochastic realization. 
   However,  much remains to be done to exploit the new approach to other multi-user problems of information theory.

\newpage
\begin{footnotesize}
\bibliographystyle{IEEEtran}
\bibliography{bibliography_paper1_new}
\bibliographystyle{plain}
\end{footnotesize}


\end{document}